\documentclass[twocolumn,preprintnumbers,amsmath,amssymb,showkeys]{revtex4}
\usepackage{amsfonts}
\usepackage{amsmath}
\usepackage{graphicx}
\usepackage{dcolumn}
\usepackage{bm}
\usepackage{overpic}
\usepackage{booktabs}
\usepackage{color}
\usepackage{multirow}

\begin{document}
\title{Reconstructing propagation networks with temporal similarity metrics}
\author{Hao Liao and An Zeng\footnote{an.zeng@unifr.ch}}

\affiliation{Department of Physics, University of Fribourg, Chemin du Mus\'{e}e 3, CH-1700 Fribourg, Switzerland}

\begin{abstract}
Node similarity is a significant property driving the growth of real networks. In this paper, based on the observed spreading results we apply the node similarity metrics to reconstruct propagation networks. We find that the reconstruction accuracy of the similarity metrics is strongly influenced by the infection rate of the spreading process. Moreover, there is a range of infection rate in which the reconstruction accuracy of some similarity metrics drops to nearly zero. In order to improve the similarity-based reconstruction method, we finally propose a temporal similarity metric to take into account the time information of the spreading. The reconstruction results are remarkably improved with the new method.
\end{abstract}
\maketitle

\section{Introduction}
One of the key features in complex networks is the similarity between nodes \cite{siam}. An accurate estimation of node similarity is related to many applications in network science, ranging from, for instance, link prediction \cite{LP} to personalized recommendation \cite{PNAS}, spurious link identification \cite{SpuriousPNAS,An2014} to backbone extraction \cite{backbone,Hub}, community detection \cite{Community,Community2} to network coarse graining \cite{CoasePRL,CoasePRE}. However, how to objectively estimate the similarity between nodes still remains a challenge in which the optimal solution depends significantly on the problems we are facing. For example, in recommender systems it has already been pointed out that a more effective similarity metric should be biased to small degree nodes to enhance diversity of the recommendation. For the problem of spurious link identification \cite{SpuriousPNAS}, the similarity metric should be combined with the betweenness index to avoid removing the important links connecting communities \cite{PRE}. The similarity is even shown to drive the network evolution together with the preferential attachment mechanism \cite{EPLPA}.

Recently, another fundamental problem attracts increasing attention: reconstructing propagation networks from observed spreading results. The spreading, as an important dynamics in networks, has been applied to simulate many real processes including epidemic contagion \cite{Epidemic,Brain}, cascading failure \cite{CatasPNAS}, rumor propagation \cite{RumorACM} and so on. In real systems, normally some partial data of the spreading process are available, but the underlying structure of the propagation network is not accessible. Therefore, how to infer the propagation networks from the collected spreading data becomes an outstanding problem. Solving this problem may help us reveal the unknown topology of many real networks, such as the terrorists' social networks \cite{TerrScience} and some biological networks which cannot be directly observed by lab instruments \cite{BioNature}.

In the literature, several works have already been done in this direction. Very recently, the compressed sensing theory has been introduced to infer the propagation networks \cite{ReconstructNature}. This technique, though effective, has relatively high computational complexity which prevents its application in large scale networks. For real networks, especially in the online social systems, the networks can contain millions of users. An efficient algorithm should be based on only local information. To solve this problem, some local similarity metrics have been applied to inferring the propagation networks \cite{An2013}. The basic idea is that nodes receiving similar information/virus in spreading are more likely to be connected in the propagation network. However, the similarity-based methods only use the final spreading results as input information. In reality, one may be able to access more detailed spreading information even including the time stamp that records when the information/virus reaches the node. Such information, if used properly, may significantly improve the inference accuracy.

Even though there are many problems, such as link prediction \cite{LP} and personalized recommendation \cite{PNAS}, related to the network reconstruction, they are essentially different. In link prediction and personalized recommendation, the main task is to estimate the likelihood of a nonexisting link to be a real link in the future \cite{LP}. A method that can place more real links on the top of the likelihood ranking has high accuracy. In network reconstruction, the accuracy is not the only focus. A well-performed method should also avoid high ranking of the false links that may result in significant difference between the reconstructed network and the real network. Therefore, one may reach completely different conclusions even if the same similarity method is applied to these two different types of problems (As an example, see \cite{PRE}). In this context, the performance of the existing similarity metrics has to be reexamined when applied to network reconstruction.

In this paper, we first systematically studied the performance of different similarity metrics which used for reconstructing the propagation networks. Interestingly, some methods, which generally enjoy high accuracy in predicting missing links, perform very badly in reconstructing the propagation networks under some infection rates. We find that this is because these similarity metrics overwhelmingly suppress large degree nodes, so that the links are mostly connected to the original small degree nodes. Moreover, we find a phenomenon called "too much information equals to no information": when the infection rate is higher than the critical value, each information/virus will cover a large part of the network, making the similarity metric fail to capture the local structure of the network. In order to solve this problem, we propose a temporal similarity metric to incorporate the time information of the spreading results. The simulation results in both artificial and real networks show that the reconstruction accuracy is remarkably improved with the new method.

\section{Model}
In this paper, we make use of the well-known Susceptible-Infected-Remove (SIR) model to simulate the spreading process on networks \cite{RMP}. Although it is an epidemic spreading model, it has also been applied to model the information propagation process \cite{RumorSIR}. We here use the news propagation as an example, but we remark that our method can also be applied to the epidemic spreading case.

A social network with $N$ nodes and $E$ links can be represented by an adjacency matrix $A$, with $A_{ij}=1$ if there is a link between node $i$ and $j$, and $A_{ij}=0$ otherwise. In our model, each node has a probability $f$ submitting a piece of news to the network. As there are $N$ nodes in the network, finally there will be $f\times N$ pieces of news propagating in the networks. The propagation of the news follows the rule of the SIR model: After the news/story $\alpha$ is submitted (or received) by a node, it will infect each of this node's susceptible neighbors with probability $\beta$. After infecting neighbors, the node will immediately be marked as recovered. During the spreading, we record all the news that each node receives. Moreover, the time step that the news was received by each node is also recorded. At the end, the information of news received by nodes is stored in an matrix $R$, with $R_{i\alpha}=1$ if $i$ have received news $\alpha$, and $R_{i\alpha}=0$ otherwise. When $R_{i\alpha}=1$, the time step at which $i$ received $\alpha$ is recorded in $T_{a\alpha}$. In this way, the temporal information of the news propagation is all stored in matrix $T$. The main task is to use the information of $R$ and $T$ to rebuild the network $A$.

\section{Methods}
The methods we used to reconstruct the network will be based on node similarity. The basic idea is that the nodes receiving many common news are similar and tend to link together in the networks. Therefore, the similarity $s_{ij}$ between node pair $ij$ can be used to estimate the likelihood $L_{ij}$ for two nodes to have a link in the network. With $R$, many similarity methods can be used to calculate the similarity between nodes. The performance of these methods have been extensively investigated in ref. \cite{LuZLinkPred,SimNature}. Here, we mainly consider four representative methods: Common neighbors \cite{siam}, Jaccard \cite{jaccard}, Resource Allocation \cite{RA} and Leicht-Holme-Newman Indices \cite{LHN}.

As we are able to get access to the information of the time step $T_{i\alpha}$ at which the news $\alpha$ are received by the node $i$, we can further improve the similarity with $T_{i\alpha}$. If two nodes receive the news at a closer time step, they are more likely to be connected in the network. Therefore, for each similarity method, we will design an improved method based on the temporal information of the news propagation. The original similarity methods and the improved ones are listed below.

(i) \emph{Common Neighbours (CN)} The common neighbor index is the simplest one to measure node similarity by directly counting the overlap of news received, namely
\begin{equation}
s_{ij}=\sum_{\alpha}R_{i\alpha}R_{j\alpha}.
\end{equation}

(ii) \emph{Temporal Common Neighbours (TCN)} This method, based on the common neighbor index, takes into account the time steps difference between two nodes receiving the news in common. The formula reads
\begin{equation}
s_{ij}=\sum_{\alpha}\frac{R_{i\alpha}R_{j\alpha}}{T_{i\alpha}-T_{j\alpha}}.
\end{equation}

(iii) \emph{Jaccard Index (Jac)} This index was proposed by Jaccard \cite{jaccard} over a hundred years ago. It can prevent the large degree nodes from having too high similarity with other nodes. The index is defined as
\begin{equation}
s_{ij} =\frac{\sum_{\alpha}R_{i\alpha}R_{j\alpha}}{\sum_{\alpha}(R_{i\alpha}+R_{j\alpha}-R_{i\alpha}R_{j\alpha})}
\end{equation}

(iv) \emph{Temporal Jaccrad Index (TJac)} The Jaccard index can also be improved by $T_{i\alpha}$ as
\begin{equation}
s_{ij} =\frac{\sum_{\alpha}R_{i\alpha}R_{j\alpha}(T_{i\alpha}-T_{j\alpha})^{-1}}{\sum_{\alpha}(R_{i\alpha}+R_{j\alpha}-R_{i\alpha}R_{j\alpha})}
\end{equation}

(v) \emph{Resource Allocation Index (RA)} The similarity between $i$ and
$j$ is defined as the amount of resource $j$ received from
$i$ \cite{RA}, which is
\begin{equation}
s_{ij} =\sum_{\alpha}\frac{R_{i\alpha}R_{j\alpha}}{\sum_{i}R_{i\alpha}}.
\end{equation}

(vi) \emph{Temporal Resource Allocation Index (TRA)} The improved RA method reads
\begin{equation}
s_{ij} =\sum_{\alpha}\frac{R_{i\alpha}R_{j\alpha}}{(T_{i\alpha}-T_{j\alpha})\sum_{i}R_{i\alpha}}.
\end{equation}

(vii) \emph{Leicht-Holme-Newman Index (LHN)} This index assigns
high similarity to node pairs that have many common
neighbours compared to the expected number of such neighbours \cite{LHN}. It is defined as
\begin{equation}
s_{ij} =\frac{\sum_{\alpha}R_{i\alpha}R_{j\alpha}}{\sum_{\alpha}R_{i\alpha}\sum_{\alpha}R_{j\alpha}}
\end{equation}

(viii) \emph{Temporal Leicht-Holme-Newman Index (TLHN)} Similar to the above three improved methods, the formula is
\begin{equation}
s_{ij} =\frac{\sum_{\alpha}R_{i\alpha}R_{j\alpha}(T_{i\alpha}-T_{j\alpha})^{-1}}{\sum_{\alpha}R_{i\alpha}\sum_{\alpha}R_{j\alpha}}.
\end{equation}

In all the temporal methods above, we set $(T_{i\alpha}-T_{j\alpha})^{-1}=0$ when $T_{i\alpha}=T_{j\alpha}$. In this case, $i$ is definitely not the node that passes the news to $j$, so $i$ and $j$ are unlikely to be connected in the networks.

\section{Metrics}
In this paper, we adopt three metrics to evaluate the performance of aforementioned methods. The first one is the standard metric of the
area under the receiver operating characteristic curve (AUC) \cite{AUC}. Each method above gives a score to all the node pairs in the network, the AUC represents the probability that a true link has a higher score than a nonexisting link. To obtain the value of the AUC,
we pick a true link and a nonexisting link in the network and compare their scores. We randomly pick up $n$ pairs of such links in total. The number of times that
the real link has a higher similarity score $s_{ij}$ than the nonexisting link is denoted as $n_1$. Moreover, we use $n_2$ to denote the number of times that the real link
and the nonexisting link have the same score $s_{ij}$. Then the AUC value is calculated as follows:
\begin{equation}
AUC=(n_1+0.5*n_2)/n
\end{equation}
Note that, if links were ranked at random, the AUC value would be equal to $0.5$. In this paper, we set $n = 10^5$.

The second and third metrics require the reconstruction of the network. The node pairs are ranked in descending order according to $s_{ij}$, and $E$ (we assume that we know roughly the number of real links in the network) top-ranked links are used to reconstruct the network. Naturally, the precision of the reconstruction, as the second metric, can be assessed by the overlap of the links in the reconstructed network and the real network. The precision metric can be regarded as the complementary measurement to AUC. The third metric is the Pearson correlation between node degree in the reconstructed network and the real network. In fact, AUC and precision measure the performance of the methods in individual level, i.e. whether the top-ranked link exist or not in the network. The degree correlation, on the other hand, evaluate the methods in rather collective level, i.e. whether the methods can correctly infer the degree of nodes.

\begin{figure*}
  \centering
  \includegraphics[width=17cm,scale=0.5]{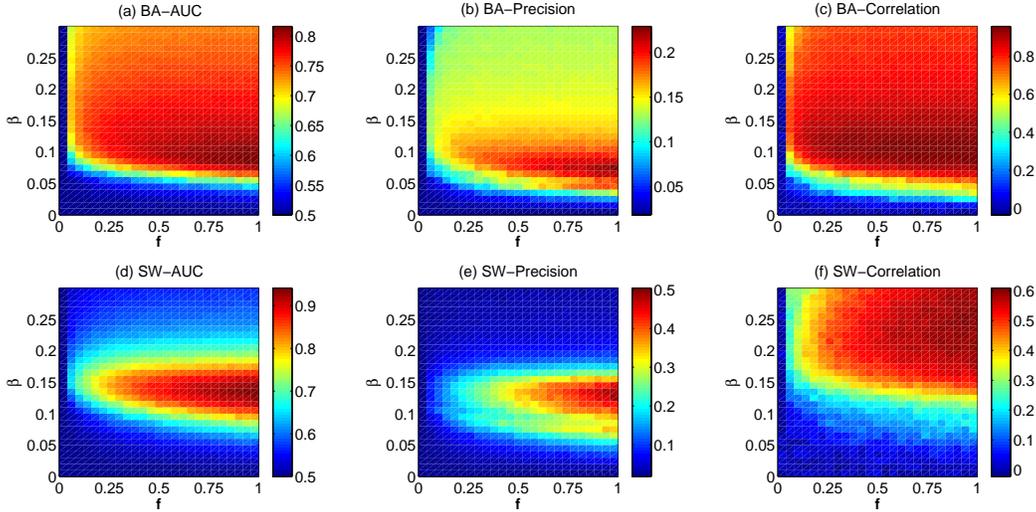}
\caption{(Color online) The $AUC$,$Precision$ and $Correlation$ in the parameter space ($\beta$, $f$) for (a,b,c) BA networks ($N=500$, $\langle k\rangle=10$)and (d,e,f) SW networks ($N=500$, $p=0.1$, $\langle k\rangle=10$) by using $CN$ method. The results are averaged over $50$ independent realizations.}
\label{fig1}
\end{figure*}

\section{Artificial networks}
We first analyze the methods in two classic artificial networks: (i) Small-World networks (SW), also known as the Watts-Strogatz model~\cite{SW}, (ii) Scale-free networks, generated by the Barabasi-Albert model (BA)~\cite{BA}. The spreading process has two parameters: infection rate $\beta$ and news submission probability $f$. With the Common Neighbor (CN) method as an example (see the results of other methods in Fig. s1, Fig. s2 and Fig. s3 in the SI), we study the influence of these two parameters on the network reconstruction results in Fig. 1. The $AUC$, precision and degree correlation in the parameter space ($\beta$, $f$) for both BA and SW networks are shown. One can see that in each panel $\beta$ significantly affects the results. In BA networks, the optimal $\beta$ resulted in the highest $AUC$, and precision and degree correlation are nearly the same (around $0.1$). However, in SW networks the optimal $\beta$ for $AUC$ and precision is different from the optimal $\beta$ for degree correlation. More specifically, to achieve the highest $AUC$ and precision, $\beta$ in SW needs to be around $0.15$. However, the best $\beta$ for degree correlation is around $0.25$. In ref.\cite{An2013}, it has already been pointed out that the optimal $\beta$ for AUC is roughly equal to $1/\langle k\rangle$. Different from $\beta$, the effect of $f$ on the results is monotonous. All the three metrics increase remarkably with $f$ when $f$ is getting small. After $f$ is higher than a threshold, these three metrics are affected only slightly by $f$.

\begin{figure*}
  \centering
  \includegraphics[width=17cm,scale=0.5]{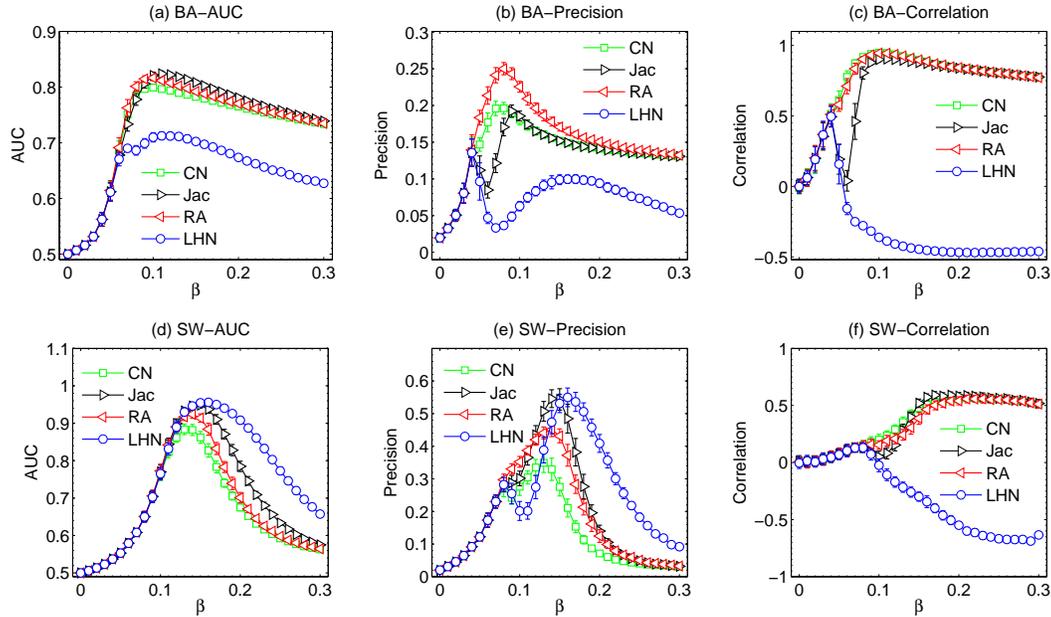}
\caption{(Color online) The dependence of the $AUC$,$Precision$ and $Correlation$ on $\beta$ with four different similarity methods in BA networks ($N=500$, $\langle k\rangle=10$)and (d,e,f) SW networks ($N=500$, $p=0.1$, $\langle k\rangle=10$). We pick $f=0.5$ here. The results are averaged over $50$ independent realizations.}
\label{fig2}
\end{figure*}

We move to compare the performance of different similarity methods. To this end, we present the dependence of $AUC$, precision and degree correlation on $\beta$ of CN, Jac, RA and LHN methods in Fig. 2 (see Fig. s4 in the SI for the dependence of the three metrics on $f$). In this figure, $f$ is set as $0.5$. As we discussed in Fig. 1, when CN is applied, one can observe a pronounced peak when tuning $\beta$. The reason for this peak has already been explained in ref. \cite{An2013}. Here, the interesting phenomenon happens when different similarity methods are compared. For Jac and LHN, the peaks in AUC still exist. However, when precision and degree correlation are considered, the curves of these two metrics drop suddenly within a certain range of $\beta$ which we refer to as the special range of $\beta$. This phenomenon can be explained by analyzing the formulae of the two similarity measurements. In Jac and LHN, the similarity between nodes are not only based on the common news these two nodes received. The overlap of news is normalized by a factor as a function of the number of news these two nodes received. The normalization is used to enhance the similarity score of the nodes receiving only a small number of news, meanwhile suppressing the similarity score of the nodes receiving many news. In the special range of $\beta$, the number of news received by large degree nodes and small degree nodes becomes remarkably different. Therefore, the normalization in Jac and LHN penalizes the large degree nodes so much that it finally gains very few links in the network reconstruction. Therefore, the Precision drops substantially and the degree correlation becomes almost zero in the special range of $\beta$. Interestingly, such decreasing of accuracy cannot be observed by the $AUC$ metric, which indicates the importance of network reconstruction when different similarity methods are evaluated. Considering the RA method outperforms CN in AUC and accuracy, we conclude here that RA is the most accurate and reliable similarity method for network reconstruction.

We use Fig. 3 to confirm our explanation above. We first pick up all the node pairs receiving at least one common piece of news. The total number of news received by each node pair $ij$ is computed and denoted as $d_{ij}$. If $d$ is homogeneously distributed, the normalization terms in Jac and LHN affect only slightly on the final similarity score. If the distribution of $d$ is overly heterogenous, some nodes with small numbers of received news will dominate the similarity score. To measure the unevenness of the distribution of $d$, we make use of the well-known Gini coefficient \cite{Giniindex}. The value of Gini is within 0 and 1. A higher Gini corresponds to a more heterogeneous distribution. In Fig. 3, we report the influence of the infection rate $\beta$ on the Gini coefficient of $d$. One can see that the standard deviation of $d$ indeed reaches an maximum when $\beta$ is in the special range.

\begin{figure}
  \centering
  \includegraphics[width=9cm,scale=0.5]{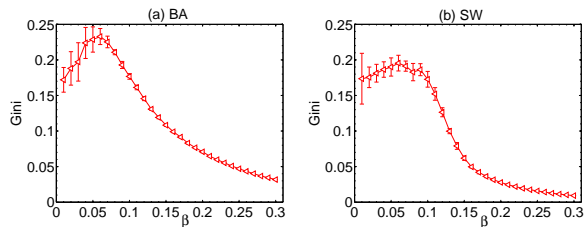}
\caption {(Color online) The dependence of the $Gini$ on $\beta$ in BA networks ($N=500$, $\langle k\rangle=10$)and SW networks ($N=500$, $p=0.1$, $\langle k\rangle=10$). Here $f = 0.5$. The results are averaged over $50$ independent realizations.}
\vspace{0.2cm}
\label{fig3}
\end{figure}

During the news propagation process, the time stamp when the news reaches each node is recorded. We thus used the temporal information of the news propagation to improve the existing similarity methods (see the Methods section). Here, we present the advantage of these temporal similarity methods in Fig. 4 and Fig. 5. In Fig. 4, we show the dependence of the AUC on $f$ and $\beta$. In Fig. 4(a), $\beta= 1\langle k\rangle$ and one can see that TCN and TJac can significantly outperform CN and Jac, respectively (see the results of other temporal similarity methods in Fig. s5 in the SI). In Fig. 4(c), $\beta= 1\langle k\rangle$ again, but the curves of the original similarity methods and the temporal similarity methods overlap, indicating the received news under this $\beta$ dominates the similarity. In Fig. (b)and (d), one interesting feature of the temporal similarity methods can be observed. When $\beta$ is large, the $AUC$ of the classic similarity methods are very low. This is because the news proposed by every node can reach a large part of the networks, so that the news coverage can no longer reflect the topology information of the network. This can be referred to as too much information equals to no information. However, when TCN and TJac methods are applied, AUC can be remain close to $1$ even when $\beta$ is as large as $0.1$. These results indicates that the temporal information is crucial to the network reconstruction from the propagation process. However, we have to remark that, when $\beta$ is small, as we see in the Fig. 4, the temporal information cannot improve the $AUC$.

\begin{figure}
  \centering
  \includegraphics[width=9cm,scale=0.5]{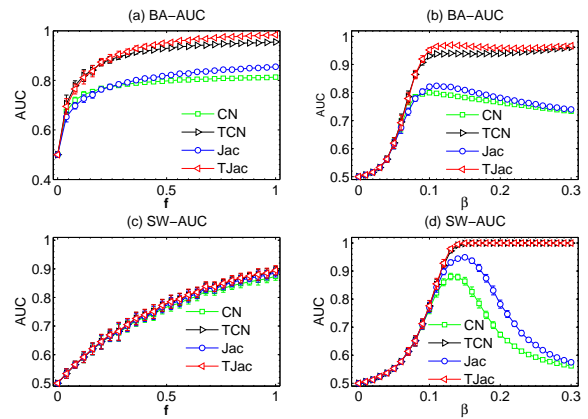}
\caption{(Color online) The dependence of the $AUC$ on $f$ with time-based similarity methods in BA and SW networks showed in (a) and (c) with setting $\beta= 1/\langle k\rangle$, and the dependence of the $AUC$ on $\beta$ for BA and SW networks with setting $f=0.5$ here. The results are averaged over $50$ independent realizations.}
\label{fig4}
\end{figure}

In Fig. 5, we study the dependence of degree correlation on $f$ and $\beta$ respectively when the temporal similarity methods are used. Clearly, the temporal similarity methods cannot improve the correlation and the special range of $\beta$ still exists. This is easy to understand as the degree correlation is mainly determined by the normalization factor of the similarity methods. Therefore, when selecting the temporal similarity method, one still needs to be very careful, as an inappropriate method may still result in a negative degree correlation and very low reconstruction accuracy. In general, the best method is the TRA method (see Fig. s5 in the SI for its performance).

\begin{figure}
  \centering
  \includegraphics[width=9cm,scale=0.5]{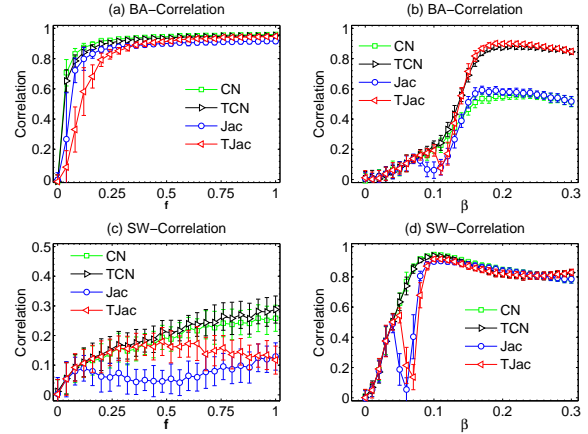}
\caption{(Color online) The dependence of the $Correlation$ on $f$ with time-based similarity methods in BA and SW networks showed in (a) and (c) with setting $\beta= 1/\langle k\rangle$, and the dependence of the $Correlation$ on $\beta$ for BA and SW networks with setting $f=0.5$ here. The results are averaged over $50$ independent realizations.}
\label{fig5}
\end{figure}

\begin{table*}
  \centering
  \caption{Basic properties of real undirected networks and the performance of the CN, TCN, Jac and TJac methods on these networks. The parameters are set as $\beta=2/\langle k\rangle$ and $f=0.5$. The similarity method with the best performance in each network is highlighted in bold font.}\label{tab1}
\begin{tabular}{p{1.2cm}|p{1.0cm}p{1cm}|p{1.0cm}p{1.0cm}p{1.0cm}p{1cm}|p{1.0cm}p{1.0cm}p{1.0cm}p{1cm}|p{1.0cm}p{1.0cm}p{1.0cm}p{1.0cm}}
\hline
\multirow{2}{*}{Network} &
\multicolumn{2}{c|}{Basic properties} &
\multicolumn{4}{c|}{AUC} &
\multicolumn{4}{c|}{Precision} &
\multicolumn{4}{c}{Correlation} \\
\cline{2-15}

 &N &E  &CN	 &	TCN	 &	Jac &	TJac	 &	CN	 &	TCN	 &	Jac	 &	TJac	 &	CN	 &	TCN	 &	Jac	 &	TJac\\	
 \hline
  Dolphins  &	62	 &	159	 &	0.779	&	0.956	&	0.826	&	\textbf{0.971}	 &	0.335	&	0.657	&	0.377	&	\textbf{0.737}	 &	0.657	&	0.764	&	0.698	&	\textbf{0.841}	 \\
  Word     &	112	 &	425	 &	0.795	&	0.916	&	0.800	&	\textbf{0.926}	 &	0.301	&	0.537	&	0.305	&	\textbf{0.551}	 &	0.761	&	0.815	&	0.758	&	\textbf{0.816}	 \\
  Jazz     &	198	 &	2742	 &	0.785	&	0.856	&	0.788	&	\textbf{0.861}	 &	0.414	&	0.521	&	0.415	&	\textbf{0.526}	 &	\textbf{0.853}	&	0.821	&	0.850	&	0.819	 \\
  E. coli   &	230	 &	695	 &	0.867	&	0.940	&	0.893	&	\textbf{0.967}	 &	0.324	&	0.524	&	0.326	&	\textbf{0.532}	 &	0.828	&	0.789	&	\textbf{0.830}	&	0.790	\\
  USAir   &	332	 &	2126	 &	0.906	&	0.928	&	0.912	&	\textbf{0.938}	 &	0.520	&	0.502	&	\textbf{0.509}	&	0.501	 &	0.820	&	\textbf{0.837}	&	0.821	&	0.836	\\
  Netsci  &	379	 &	914	 &	0.858	&	0.979	&	0.968	&	\textbf{0.998}	 &	0.213	&	0.609	&	0.443	&	\textbf{0.837}	 &	0.498	&	0.630	&	0.642	&	\textbf{0.876}	\\
  Email   &	1133	 &	5451	 &	0.828	&	0.920	&	0.834	&	\textbf{0.933}	 &	0.109	&	0.393	&	0.109	&	\textbf{0.397}	 &	0.779	&	0.851	&	0.779	&	\textbf{0.852}	 \\
  TAP   &	1373	 &	6833	 &	0.816	&	0.934	&	0.887	&	\textbf{0.990}	 &	0.175	&	0.547	&	0.261	&	\textbf{0.575}	 &	0.687	&	0.757	&	0.748	&	\textbf{0.782}	\\
  PPI   &	2375	  &	11693	 &	0.890	&	0.942	&	0.924	&	\textbf{0.972}	 &	0.289	&	0.338	&	0.289	&	\textbf{0.351}	 &	\textbf{0.792}	&	0.748	&	0.791	&	0.751	\\
\hline
\end{tabular}
\end{table*}

\begin{table*}
  \centering
  \caption{Basic properties of real directed networks and the performance of the CN, TCN, Jac and TJac methods on these networks.  The parameters are set as $\beta=2/\langle k\rangle$ and $f=0.5$. The similarity method with the best performance in each network is highlighted in bold font.}\label{tab2}
\begin{tabular}{p{1.3cm}|p{1.0cm}p{1cm}|p{1.0cm}p{1.0cm}p{1.0cm}p{1cm}|p{1.0cm}p{1.0cm}p{1.0cm}p{1cm}|p{1.0cm}p{1.0cm}p{1.0cm}p{1.0cm}}
\hline
\multirow{2}{*}{Networks} &
\multicolumn{2}{c|}{Basic properties} &
\multicolumn{4}{c|}{AUC} &
\multicolumn{4}{c|}{Precision} &
\multicolumn{4}{c}{Correlation} \\
\cline{2-15}
&N &E  &CN	 &	TCN	 &	Jac &	TJac	 &	CN	 &	TCN	 &	Jac	 &	TJac	 &	CN	 &	TCN	 &	Jac	 &	TJac\\	
 \hline
Prisoners	 &	67	 &	182	 &	0.724	 &	0.811	 &	0.797	 &	\textbf{0.841}	 &	0.215	 &	0.468	 &	0.411	 &	\textbf{0.575}	 &	0.573	 &	0.686	 &	0.677	 &	\textbf{0.730}	 \\
SM FW	 &	54	 &	356	 &	0.646	 &	\textbf{0.666}	 &	0.634	 &	0.662	 &	0.255	 &	\textbf{0.287}	 &	0.236	 &	0.283	 &	0.869	 &	\textbf{0.885}	 &	0.754	 &	0.843	 \\
Neural	 &	297	 &	2359	 &	0.722	 &	0.794	 &	0.731	 &	\textbf{0.809}	 &	0.144	 &	0.251	 &	0.143	 &	\textbf{0.290}	 &	\textbf{0.684}	 &	0.592	 &	0.553	 &	0.512	 \\
Metabolic	 &	453	 &	2040	 &	0.683	 &	0.703	 &	0.703	 &	\textbf{0.722}	 &	0.088	 &	0.135	 &	0.140	 &	\textbf{0.228}	 &	0.541	 &	0.640	 &	0.598	 &	\textbf{0.715}	 \\
PB	 &	1222	 &	19090	 &	0.844	 &	0.860	 &	0.844	 &	\textbf{0.861}	 &	0.151	 &	0.251	 &	0.155	 &	\textbf{0.252}	 &	\textbf{0.809}	 &	0.803	 &	0.800	 &	0.802	 \\
\hline
\end{tabular}
\end{table*}

\section{Real undirected networks}
We further apply the methods on the real networks. Firstly, the methods are applied to real undirected networks. We consider nine empirical networks including both social networks and nonsocial networks: (i) Dolphin: an undirected social network of frequent associations between 62 dolphins in a community living off Doubtful Sound, New Zealand \cite{dolphins}. (ii) Word: adjacency network of common adjectives and nouns in the novel David Copperfield written by Charles Dickens \cite{netcoauthor_word}. (iii) Jazz: a music collaboration network obtained from the Red Hot Jazz Archive digital database. It includes 198 bands that performed between 1912 and 1940, with most of the bands from $1920$ to $1940$ \cite{jazz}. (iv) E.coli: the metabolic network of E.coli \cite{Ecoli}. (v) USAir: the US air transportation network \cite{USAir}. (vi) Netsci: a coauthorship network between scientists who published on the topic of network science \cite{netcoauthor_word}. (vii) Email: an email communication network~\cite{email}. (viii) TAP: a yeast protein binding network generated by tandem affinity purification experiments \cite{TAP}. (ix) PPI: a protein-protein interaction network~\cite{PPI}. We only take into account the giant component of these networks. This is because a pair of nodes located in two disconnected components, their $s_{xy}$ score will be zero according to CN and its variant.

The results of the similarity methods on these networks are detailedly reported in Table 1. Consistent with the results in the artificial networks, the temporal similarity methods significantly outperforms the classic similarity methods (not necessarily in degree correlation). In Table 1, TJac outperforms TCN in both AUC and Precision. The results of TLHN and TRA methods are reported in Table s1. The special zone is also observed when LHN methods is applied to real networks. For example, in the email network, the degree correlation drops to negative when $\beta>0.1$, and the precision value is significantly lowered (from 0.2 to 0.02). However, we also observe that Jac no longer leads to the sudden drop of correlation and precision in the real networks we considered. Comparing all the methods, the TRA method in generally enjoys the highest accuracy.

\section{Real directed networks}
The methods are also applied to real directed networks. We considered several real directed networks to validate our methods. Results of TCN and TJac are shown in table 2 and results of TLHN and TRA methods are shown in Table s2.  The networks include Prisoners (friendship network between prisoners) \cite{Prision}, St. Marks FW (food web in St. Mark area) \cite{FW}, C. elegans neural (neural network of C. elegans) \cite{Celegans}, C. elegans metabolic (metabolic network of C. elegans) \cite{Celegans}, PB (hyper link between the blogs of politicians) \cite{PB}.

Like the undirected networks, the temporal similarity methods have a much higher AUC and precision than the classic similarity methods. However, one can also see that AUC and Precision in directed networks are on average lower than the undirected networks. This indicates that it is generally more difficult to reconstruct directed networks via similarity metrics. We also studied the effect of $\beta$ on the results in directed networks. We observe that the improvement of the temporal similarity methods becomes more significant when $\beta$ is larger. Moreover, the special zone of both the Jac and LHN methods exists when adjusting $\beta$ in directed networks. Taking the Neural network as an example, when LHN is applied and $\beta>0.08$, the degree correlation drops to negative and the precision decreases from 0.15 to 0.07. We remark that the results on other networks are similar.

\section{Discussion}
In this paper, we applied some standard similarity metrics to reconstruct the propagation network based on the observed spreading results. We find that even though some similarity methods such as Jaccard and LHN perform well in link prediction, they may cause some serious problem when it is used to reconstruct networks, as they may assign many links to the nodes that suppose to have low degree. We find that the resource allocation method not only has high reconstruction accuracy, but also results in similar network structural properties as the real network. Finally, we take into account the temporal information of the propagation process, and we find that such information can significantly improve the reconstruction accuracy of the existing similarity methods, especially when the infection rate is large.

Some problems still remain unsolved. For example, our methods now requires the full time information. When only partial time information is available, the temporal similarity methods should be modified. In addition our work only consider the simplest epidemic spreading model. Other more realistic models describing the disease contagion and information propagation needs to be examined \cite{PR}. Furthermore, many similar problems in other fields also needs to be addressed. For instance, most link prediction methods are based on the observed network topology. When the time information of the observed links is available, the similarity methods should be modified accordingly to incorporate the temporal information of the network. The node similarity is the basic feature for community detection. Improving the detection accuracy with the time information would be an important task. We believe our work may inspire some solution to the above problems in the near future.

\textbf{Acknowledgement.} We thank Prof.Yi-Cheng Zhang for fruitful discussion. This work was partially supported by the EU FP7 Grant 611272 (project GROWTHCOM) and by the Swiss National Science Foundation (grant no.~200020-143272). The author would like to acknowledge the support from China Scholarship Council.

\end{document}